\documentclass[twocolumn,aps,prd,showpacs,nofootinbib,floatfix,eqsecnum]{revtex4}  
\usepackage{amsmath,latexsym,graphicx,multirow,longtable}     
\begin{document}  
\title{Towards adiabatic waveforms for inspiral into Kerr black  
holes:\\ II. Dynamical sources and generic orbits}  
\author{Pranesh A.\ Sundararajan${}^{1}$}  
\author{Gaurav Khanna${}^{2}$}  
\author{Scott A.\ Hughes${}^{1}$}  
\author{Steve Drasco${}^{3}$}  
  
\affiliation{${}^{1}$Department of Physics and MIT Kavli Institute, MIT, 77 Massachusetts Ave., Cambridge, MA 02139 \\ ${}^{2}$Department of Physics, University of Massachusetts, Dartmouth, MA 02747 \\ ${}^{3}$Jet Propulsion Laboratory, California Institute of Technology, Pasadena CA 91109}  
  
\date{\today}  
\begin{abstract}  
This is the second in a series of papers whose aim is to generate ``adiabatic'' gravitational waveforms from the inspiral of stellar-mass compact objects into massive black holes.  In earlier 
work, we presented an accurate (2+1)D finite-difference time-domain 
code to solve the Teukolsky equation, which evolves curvature 
perturbations near rotating (Kerr) black holes.  The key new 
ingredient there was a simple but accurate model of the singular  
source term based on a discrete representation of the Dirac-delta  
function and its derivatives.  Our earlier work was intended as a  
proof of concept, using simple circular, equatorial geodesic orbits as  
a testbed.  Such a source is effectively static, in that the smaller  
body remains at the same coordinate radius and orbital inclination  
over an orbit.  (It of course moves through axial angle, but we  
separate that degree of freedom from the problem.  Our numerical grid  
has only radial, polar, and time coordinates.)  We now extend the  
time-domain code so that it can accommodate dynamic sources that move  
on a variety of physically interesting world lines.  We validate the  
code with extensive comparison to frequency-domain waveforms for cases  
in which the source moves along generic (inclined and eccentric) bound  
geodesic orbits.  We also demonstrate the ability of the time-domain  
code to accommodate sources moving on interesting non-geodesic  
worldlines.  We do this by computing the waveform produced by a test  
mass following a ``kludged'' inspiral trajectory, made of bound  
geodesic segments driven toward merger by an approximate radiation  
loss formula.  
\end{abstract}  
\maketitle  
  
\section{Introduction}  
\label{sec:intro}  
  
\subsection{Background}  
\label{sec:background}  
  
The extreme mass ratio limit of general relativity's two-body problem  
has been a major focus of work in recent years.  This limit  
corresponds to a stellar mass compact object that orbits and perturbs  
a massive black hole.  The system generates gravitational waves (GWs)  
which drive the small body to inspiral into the large black hole.  
Measuring such ``extreme mass ratio inspiral,'' or {\it EMRI}, events  
is a major goal for space-based GW antennae, particularly the LISA  
mission\footnote{http://lisa.nasa.gov, http://sci.esa.int/lisa}.  
EMRIs should be measurable to a redshift $z \sim 0.5 - 1$.  The event  
rate at this range is estimated to be high enough that a multiyear  
LISA mission should measure dozens to hundreds of EMRI events  
{\cite{getal04}}.  Because the smaller body only slightly perturbs the  
larger black hole's spacetime, EMRI GWs are expected to provide an  
exceptionally clean probe of black hole properties.  We expect to use  
EMRIs to measure black hole masses and spins with extremely good  
accuracy {\cite{bc04}}, and even to test how well the spacetime meets  
the rather stringent constraints that the ``no-hair'' theorems of  
general relativity impose on black holes  
{\cite{ch04,gb06,bc07,glm08}}.  
  
Understanding EMRI sources will require us to compare measured waves  
with theoretical models that are as accurate as possible.  This goal  
motivates much recent EMRI work.  The waves are sufficiently  
complicated that simply {\it detecting} them in LISA's datastream will  
be a challenge.  Techniques for finding these events are currently  
being developed and tested through the ``Mock LISA Data Challenges'',  
or MLDCs (see Refs.\ {\cite{amaldi_mldc,vallis_gwdaw}} for overviews  
of recent MLDCs).  An important input to these challenges (and to the  
development of EMRI measurement techniques more generally) are  
waveform models that capture the true complexity of EMRI events (see  
{\cite{cornish_gwdaw,gair_gwdaw}} for discussion of recent work to  
include EMRI waves in the MLDCs).  
  
This paper presents a further step in our program to construct  
accurate EMRI wave models.  As discussed in the introduction to Ref.\  
{\cite{skh07}} (Paper I), our goal is to make ``adiabatic'' waveforms  
--- waveforms built by separately treating the long-time dissipative  
evolution and the short-time conservative motion.  In our present  
analysis, we take the short-time motion to be a geodesic orbit of the  
background spacetime; our approach thus amounts to approximating the  
inspiral trajectory as a sequence of geodesic orbits.  As discussed by  
Pound and Poisson {\cite{pp08}}, this limit is more properly a  
``radiative'' or ``dissipative'' approximation, since we do not  
include conservative self interactions.  It may be possible to augment  
this analysis with at least some conservative effects {\cite{jf}}, so  
we believe the program we are developing is capable of building truly  
adiabatic inspiral waveforms as described in {\cite{pp08}}.  We will  
describe our goal as ``adiabatic'' waveforms, but the reader should  
bear in mind that the approximation we are currently developing is  
more restricted than this.  
  
Geodesic orbits are described (up to initial conditions) by three  
conserved constants: energy $E$, axial angular momentum $L_z$, and  
``Carter constant'' $Q$.  Using black hole perturbation theory, we  
compute the rate at which these three constants evolve; fast and  
accurate frequency-domain codes make it possible to compute these  
rates of change fairly easily {\cite{dh06,sthgn06,dfh05}}.  We then  
build the parameter-space trajectory $[E(t), L_z(t), Q(t)]$ followed  
by the small body; choosing initial conditions, it is simple to build  
the coordinate-space worldline $[r(t), \theta(t), \phi(t)]$ of a  
particular inspiral.  From this worldline, we build the source to a  
time-domain code.  The output of this code is, at last, our model EMRI  
wave.  
  
\subsection{Time-domain black hole perturbation theory}  
\label{sec:td_background}  
  
Since the frequency-domain portion of this program is already well in  
hand, our current focus is on the time-domain code.  In essence, our  
goal is to build a code which takes as input any physically reasonable  
worldline, and provides as output the waveform produced by a small  
body on this worldline.  In Paper I, we demonstrated an  
accurate (2+1)D numerical code to solve, in the time domain, the wave  
equation for curvature perturbations to a black hole --- the Teukolsky  
equation {\cite{teuk}}.  Our code evolves the Weyl curvature scalar  
$\Psi_4$, constructed by projecting the vacuum curvature onto  
appropriate components of a null tetrad; see Paper I for details.  The  
azimuthal dependence of $\Psi_4$ is separated out (due to the  
$\phi$ symmetry of black holes); the dependence on the Boyer-Lindquist  
coordinates $r$, $\theta$, and $t$ is found by evolving $\Psi_4$ on a  
numerical $r$-$\theta$ grid.  
  
As is common in black hole perturbation theory, we treat the smaller  
body as a Dirac-delta point particle, leading to a singular source for  
the Teukolsky equation.  In the frequency domain, the delta can be  
dealt with analytically, and presents no great challenge.  By  
constrast, accurately computing the effect of a sharp source on the  
time-domain code's numerical grid can be extremely challenging.  In  
Paper I, we presented a new technique for treating the singular source  
term.  Our innovation was to model the delta as a series of finite  
impulses, with the largest impulse located close to the delta's  
argument, falling off rapidly as we move away from this ``central''  
spike.  Importantly, this approach allows us to accurately model the  
{\it derivatives} of the delta function.  Since the Teukolsky equation  
source depends on first and second derivatives of the delta (as well  
as the delta itself), this appears to give us an accuracy boost  
relative to other finite-difference delta representations (such as a  
truncated Gaussian), which may accurately capture the delta's  
behavior, but not do so well with the derivatives.  
  
\subsection{This paper}  
\label{sec:thispaper}  
  
Paper I focused on the properties of this new source representation.  
To clarify this focus, we studied very simple orbits: We only  
considered the (astrophysically unlikely) case of circular, equatorial  
black hole orbits.  We now extend this to include inclined, eccentric  
and generic orbits, as well as non-geodesic inspiral sequences.  
  
A particle in a circular, equatorial orbit has constant radial and  
angular coordinate, confining it to a fixed location on the $r$-$\theta$  
grid.  Eccentricity means that the orbit oscillates radially, crossing  
radial grid zones.  Similarly, orbital inclination results in angular  
grid crossing.  We quickly discovered that these new motions introduce  
high frequency numerical noise.  This noise can be controlled by  
combining a low pass filter with a higher order discretization of the  
delta function; details are given in Sec.\ {\ref{sec:delta}}.  Aside  
from this mild extension of the basic formalism presented in Paper I,  
it was not terribly difficult to use our new source term to handle a  
broad class of astrophysically interesting orbits.  To validate our  
results, we present in Sec.\ {\ref{sec:comparisons}} extensive  
comparisions with waveform snapshots computed in the frequency domain  
{\cite{dh06}}, demonstrating graphically and quantitatively (with  
appropriate overlap integrals) excellent agreement between the two  
techniques.  
  
As extensively discussed in the introduction to Paper I and here, our  
goal is to compute the waves from inspiral of a small body through a  
sequence of orbits.  As a proof-of-concept demonstration of the  
feasibility of this idea, we present a simple example of inspiral in  
Sec.\ {\ref{sec:inspiral}}.  In this example, we evolve through our  
geodesic sequence using a ``kludged'' approximation to the rates of  
change of orbital constants, using the code described in Refs.\  
{\cite{gg06,bfggh}}.  These waveforms are not reliable EMRI models,  
but they illustrate the ease with which we can handle the effect of  
radiation emission on the motion of the source.  Computing waves from  
an inspiral is no more of a computational challenge than computing  
waves from a bound geodesic.  
  
The next step will be to combine accurate radiative backreaction with  
our time-domain solver to compute ``adiabatic'' EMRI waveforms (albeit  
ones that still neglect conservative self interactions).  Plans for  
this next step are described in our final summary, Sec.\  
{\ref{sec:summary}}.  
   
\section{Dynamically varying discrete delta functions}  
\label{sec:delta}  
  
In Paper I, we presented a method for representing a Dirac delta  
function and its derivatives on a discrete numerical grid.  In that  
paper, we only considered a delta with fixed radial and angular  
position.  Naive application of the discrete delta models presented in  
Paper I leads to instabilities when the particle moves in the  
numerical grid.  The following argument outlines the root cause of  
these instabilities.  Consider the function $\delta[x - \alpha(t)]$,  
where $x_k \leq \alpha(t) \leq x_{k+1}$; i.e., the delta's peak varies  
with time and lies between two discrete grid points.  Let $x_i$  
represent any discrete point on our grid, and let $h = x_{k+1} - x_k =  
x_k - x_{k-1}$ be the grid resolution.  Naive application of the  
results from Paper I might lead us to model the delta function with  
the impulse weights  
\begin{eqnarray}  
\delta_i(t_n) & = & \frac{\alpha(t_n) - x_k}{h^2} \mbox{ for } i=k+1  \\  
    & = & \frac{x_{k+1} - \alpha(t_n)}{h^2} \mbox{ for } i=k  \\  
    & = & 0 \mbox{ everywhere else } \;.  
\end{eqnarray}  
(This ``two impulse'' delta is in fact just the simplest  
representation we developed in Paper I, but is useful for the  
following discussion.)  Each $t_n$ defines a time slice of our  
$r$-$\theta$ grid.  As $\alpha$ varies from one time slice to another,  
so do the coefficients at $x_k$ and $x_{k+1}$.  The frequency spectrum  
of $\delta_k(t_n)$ and $\delta_{k+1}(t_n)$ will reflect the amount of  
variation in $\alpha$.  A large variation in $\alpha$ will produce a  
high frequency component in the Fourier transform of the time series  
of each weight. These variations couple to the time derivatives in the  
homogeneous part of the Teukolsky equation.  Consequently, the  
solution contains spurious high frequency features of numerical  
origin.  
  
Consider the extreme limit of this effect: $\alpha$ changes so rapidly  
that the delta's peak moves across a grid zone in a single time slice:  
\begin{equation}  
\alpha(t_1) = \alpha(t_0) - h\;,  
\end{equation}  
so that  
\begin{equation}  
x_{k} \leq \alpha(t_0) \leq x_{k + 1}  
\end{equation}  
but  
\begin{equation}  
x_{k - 1} \leq \alpha(t_1) \leq x_k\;.  
\end{equation}  
The weight of the delta function very suddenly becomes zero at  
$x_{k+1}$ as we step from $t = t_0$ to $t = t_1$; likewise, the weight  
at $x_{k-1}$ very suddenly jumps from non-zero to zero in this step.  
The coupling of this sudden change to numerical time derivatives  
drives instabilities in our code, in a manner reminiscent of the  
initial burst of radiation that occurs due to the sudden appearance of  
the particle at the start of our evolution; see Fig.\ 2 of Paper I.  
  
This problem is substantially mitigated by using a delta  
representation with a wider stencil; examples of this are described in  
Paper I.  Wide stencils reduce the amount by which each weight changes  
from step to step, thereby reducing numerical noise.  Another useful  
tool is to increase the order of the delta representation, thereby  
increasing the smoothness of the delta and its derivatives.  This is  
particularly important since the Teukolsky equation is a second-order  
differential equation; some smoothness in the derivatives is necessary  
to prevent the differential operator from seeding excessive noise.  
Finally, residual high frequency noise can be removed by convolving  
the source with a low pass filter\footnote{An obvious brute force  
workaround left off this list is to simply make the grid extremely  
fine and use tiny time steps.  This does not address the root cause of  
instabilities seeded by particle motion, though it is certainly  
something used in practice (to the extent that computational limits  
allow).}.  These three techniques are each described in the following  
subsections.  
  
Each of these techniques smear out the delta function, pushing us away  
from the idealization of a zero width singularity.  Choosing between  
stability (which tends to push us to a wider delta) and faithful  
representation of the singularity (which pushes us to a narrow delta)  
leads us to an optimization problem; we tune our delta representation  
in a way that (hopefully) minimizes numerical noise and maximizes  
accuracy.  Note also that, in addition to high-frequency noise  
generated by abrupt movement of the delta across the grid, spurious  
excitations of the quasinormal modes of the black hole also appear due  
this motion.  This source of ``noise'' appears to be controlled by  
grid resolution --- wider grids lead to less pointlike deltas, which  
spuriously excite these modes.  This spurious contribution to the EMRI  
waves can be mitigated with a form of Richardson extrapolation  
{\cite{numrec}}.  We discuss this further in Sec.\  
{\ref{sec:comparisons}} and the Appendix.  
  
\subsection{Higher order delta functions}  
\label{sec:highorder}  
  
Discrete delta representations based on linear and cubic interpolation  
were derived in Paper I.  We now extend this process to arbitrary  
polynomial order, equipping us with an entire family of discrete delta  
functions.  
  
As in Paper I, we start from the defining integral,  
\begin{eqnarray}  
\int_{\alpha(t)-\epsilon}^{\alpha(t) + \epsilon}  
dx\;f(x)\;\delta[x-\alpha(t)] & = & f[\alpha(t)]\;.  
\label{eq:deltadef}  
\end{eqnarray}  
Let $x_{k+n-1} \leq \alpha \leq x_{k+n}$; the reason for our somewhat  
idiosyncratic choice of subscripts will become clear as we proceed.  
For clarity, we will not explicitly write out the time dependence of  
$\alpha$; the reader should bear in mind that $\alpha = \alpha(t)$ in  
all that follows.  Rewriting Eq.\ (\ref{eq:deltadef}) as a sum over a  
finite step size, we have  
\begin{eqnarray}  
\int_{\alpha-\epsilon}^{\alpha+\epsilon} dx\;f(x)\;\delta(x-\alpha)  
&\simeq& h\sum_{i}f\left(x_i\right)\delta_i  
\nonumber\\  
\Rightarrow f\left(\alpha\right) & \simeq &  
h\sum_{i}f\left(x_i\right)\delta_i\;.  
\label{eq:delta_used1}  
\end{eqnarray}  
  
The function $f(\alpha)$ can be approximated by the Lagrange  
interpolating polynomial,  
\begin{eqnarray}  
f(\alpha) & = & \sum_{i=k}^{k+2n-1} \frac{\Pi(\alpha)}{(\alpha-x_i) \Pi^\prime(x_i)} f(x_i)\;,   
\end{eqnarray}  
where $2n$ is the order of interpolation and  
\begin{eqnarray}  
\Pi(\alpha) &=&  \prod_{i=k}^{k+2n-1} (\alpha-x_i)  
			= \prod_{i=0}^{2n-1} (\alpha-x_{k+i})\;\; \\  
\Pi^\prime(x_j) &=& \left[\frac{d\Pi}{d\alpha}\right]_{\alpha=x_j}  
= \prod_{i=k,i\neq j}^{k+2n-1} (x_j-x_k) \;.  
\end{eqnarray}  
Inserting this in Eq.\ (\ref{eq:delta_used1}) leaves us with  
\begin{eqnarray}  
\!\!\!\!\sum_{i=k}^{k+2n-1}\!\!\frac{\Pi(\alpha)}{(\alpha-x_i)  
\Pi^\prime(x_i)}f(x_i) & = & h\sum_{i}f\left(x_i\right)\delta_i\,;  
\end{eqnarray}  
comparing coefficients of $f(x_i)$ allows us to read off $\delta_i$,  
\begin{eqnarray}  
  \delta_i & = & \frac{\Pi(\alpha)}{h(\alpha-x_i) \Pi^\prime(x_i)}.  
\label{eq:gen_delta}  
\end{eqnarray}  
We thus see that $\delta_i$ is non-zero for $i \in [k,k+2n-1]$.   
  
The weights for derivatives of the delta function can be obtained  
similarly.  Writing the identities  
\begin{eqnarray}  
\int dx\;f(x)\;\delta^\prime(x-\alpha) & =& - f^\prime(\alpha) \;\\  
\int dx\;f(x)\;\delta^{\prime\prime}(x-\alpha) & = & f^{\prime\prime}(\alpha)\;  
\end{eqnarray}   
as sums gives us  
\begin{eqnarray}  
\label{delta_identity1}  
h \sum_{i}\;f(x_i)\;\delta^\prime_i & \simeq & -f^\prime(\alpha) \nonumber\\  
	& = & -h \sum_i f'(x_i)\delta_i \nonumber \\  
\!\!\!\Rightarrow \sum_{i}\;f(x_i)\;\delta^\prime_i & = & -\!\!\!\sum_{i=k}^{k+2n-1}\!\!\! \frac{\Pi(\alpha)f^{\prime}(x_i)}{h(\alpha-x_i) \Pi^\prime(x_i)} \;,  
\end{eqnarray}  
\begin{eqnarray}  
\label{delta_identity2}  
h \sum_{i}\;f(x_i)\;\delta^{\prime\prime}_i & \simeq & f^{\prime\prime}(\alpha)\nonumber\\  
	& = & h\sum_{i}\;f^{\prime\prime}(x_i)\delta_i \nonumber \\  
\Rightarrow \sum_{i}\;f(x_i)\;\delta^{\prime\prime}_i	& = & \sum_{i=k}^{k+2n-1}\!\! \frac{\Pi(\alpha)f^{\prime\prime}(x_i)}{h(\alpha-x_i) \Pi^\prime(x_i)} \;.  
\end{eqnarray}   
We now insert centered finite difference formulae for the  
derivatives of $f(x_i)$ to obtain  
\begin{eqnarray}  
\label{eq:gen_delta_der1}  
\sum_{i}\;f(x_i)\;\delta^\prime_i & = & -\!\!\!\sum_{i=k}^{k+2n-1}\!\!\! \frac{\Pi(\alpha)}{h(\alpha-x_i) \Pi^\prime(x_i)}\times \; \nonumber \\  
&  & \left[\frac{f\left(x_{i+1}\right)-f\left(x_{i-1}\right)}{2h}\right] \;,\\   
\label{eq:gen_delta_der2}			  
\sum_{i}\;f(x_i)\;\delta^{\prime\prime}_i & = & \sum_{i=k}^{k+2n-1}\!\! \frac{\Pi(\alpha)}{h(\alpha-x_i) \Pi^\prime(x_i)} \times \nonumber\\  
			&  &\!\!\!\!\!\!\!\!\!\!\!\!\!\!\!\left[ \frac{f(x_{i+1}) - 2 f(x_{i}) + f(x_{i-1})}{h^2}\right]\!.  
\end{eqnarray}  
Expressions (\ref{eq:gen_delta_der1}) and (\ref{eq:gen_delta_der2})  
are in a form that makes it simple to read off $\delta^\prime_i$ and  
$\delta^{\prime\prime}_i$.  For example, $\delta^\prime_j$ can be  
calculated by setting $f(x_j) = 1$ and $f(x_l)=0,\; l\neq j$.  It is  
straightforward to verify that setting $n = 1$ and $n = 2$ reproduces  
the weights given by the two-point linear hat and the cubic formulae  
(described in Paper I) respectively.  We also note that the delta  
derivative coefficients are non-zero for $i \in [k - 1, k + 2n]$.  
  
\subsection{Wider stencils at a given interpolation order}  
\label{sec:widestencils}  
  
In Paper I, we generalized the two-point linear hat delta function  
such that it can be represented over a larger number of points.  
Similarly, we develop a procedure to widen the stencil of the  
generalized model obtained from Eqs.\ (\ref{eq:gen_delta}),  
(\ref{eq:gen_delta_der1}), and (\ref{eq:gen_delta_der2}).  
  
Consider a model for $\delta_i$ obtained from Eq.\  
(\ref{eq:gen_delta}) for some $n = m$. Then, $\delta_i \neq 0$ for $i  
\in [k, \ldots, k + 2m - 1]$.  Our goal is to widen this  
representation by some integer factor $w$ such that the coefficients  
are non-zero for a wider range of grid points.  Let us label the  
weights of this wider representation by $\delta^w_i$, with $\delta^w_i  
\neq 0$ for $i \in [k,...,k+2wm-1]$.  It should be emphasized that  
this is different from simply using Eq.\ (\ref{eq:gen_delta}) with $n  
= wm$; we have not changed the polynomial order, it remains fixed at  
$2m$.  
  
For concreteness, let us choose $w = 2$, doubling the number of points  
in the delta representation.  We infer the coefficients $\delta^2_i$  
at gridpoints $i = k$, $k + 2$, $k + 4$, $\ldots$, $k + 4m - 2$, by  
widening the grid by a factor of two: We evaluate $\delta_i$ with $h  
\rightarrow 2h$, $x_{k + j} \rightarrow x_{k + 2j}$ to get  
\begin{eqnarray}  
\delta^2_{k+2j} & = & \delta_{k+j} \rfloor_{h\rightarrow 2h\mbox{,} x_{k+j}\rightarrow x_{k+2j}} \nonumber\\  
& = & \frac{\Pi(\alpha)}{2h(\alpha-x_{k+2j}) \Pi^\prime(x_{k+2j})}\;,  
\end{eqnarray}  
where  
\begin{equation}  
\Pi(\alpha) = \prod_{i=0}^{2m-1} (\alpha-x_{k+2i}) \;.  
\end{equation}  
Finally, we need $\delta^2_i$ at the intermediate points $i = k + 1$,  
$k + 3$, $\ldots$, $k + 4m - 1$.  We do this by exploiting the  
translational symmetry of the problem.  Momentarily reinsert the time  
dependence of the $\delta$'s and $\alpha$.  Now consider the  
hypothetical situation where  
\begin{eqnarray}  
\alpha(t_0) = \alpha_0 \;,\nonumber\\  
\alpha(t_1) = \alpha_0 - h\;;  
\end{eqnarray}  
i.e, $\alpha(t)$ changes by a grid spacing from $t_0$ to $t_1$.  We  
must have  
\begin{eqnarray}  
\delta^2_{k+2j}(t_1) & = &\delta^2_{k+2j+1}(t_0) \nonumber\\  
\Rightarrow  
\delta^2_{k+2j}(t_0)\rfloor_{\alpha(t)\rightarrow\alpha_0-h} & = &  
\delta^2_{k+2j+1}(t_0)\rfloor_{\alpha(t)\rightarrow\alpha_0} \;.  
\end{eqnarray}   
We can turn this equation the other way around to read off the  
coefficient $\delta^2_{k+2j+1}$ at $t_0$: Simply replace $\alpha(t)$  
with $\alpha(t)-h$ in the formula for $\delta^2_{k+2j}(t_0)$ to obtain  
$\delta^2_{k+2j+1}(t_0)$.  Since there was nothing special about our  
time slice, $t_0$, we find  
\begin{eqnarray}  
\delta^2_{k+2j+1}(t_n) & = & \delta^2_{k+2j}(t_n)\rfloor_{\alpha(t)\rightarrow\alpha(t)-h}  
\end{eqnarray}  
for {\it any} moment $t_n$.  
  
Though we chose $w = 2$ for concreteness, the above argument can be  
generalized to any integer $w$.  Since our result holds for all time  
slices, we again suppress the time dependence to obtain expressions  
for any integer $w$:  
\begin{eqnarray}  
\Pi(\alpha) & = & \prod_{i=0}^{2m-1} (\alpha-x_{k+wi}) \;, \\  
\delta^w_{k+wj} & = & \delta_{k+j} \rfloor_{h\rightarrow wh\mbox{,} x_{k+j}\rightarrow x_{k+wj}} \\  
				& = & \frac{\Pi(\alpha)}{wh(\alpha-x_{k+wj}) \Pi^\prime(x_{k+wj})}\;, \\  
\delta^w_{k+wj+l} & = & \delta^2_{k+wj}\rfloor_{\alpha(t)\rightarrow\alpha(t)-lh} \nonumber\\  
				 &   & \mbox{for } l \in [1,2, \ldots, w-1] \; .  
\end{eqnarray}  
  
These techniques carry over to the derivatives as well:  
\begin{eqnarray}  
\delta^{\prime w}_{k+wj} & = & \delta^\prime_{k+j} \rfloor_{h\rightarrow wh\mbox{,} x_{k+j}\rightarrow x_{k+wj}} \\  
\delta^{\prime w}_{k+wj+l} & = & \delta^{\prime w}_{k+wj}\rfloor_{\alpha(t)\rightarrow\alpha(t)-lh}\nonumber \\  
				 &   & \mbox{for } l \in [1,2, \ldots, w-1] \; ;  
\end{eqnarray}		  
and  
\begin{eqnarray}  
\delta^{\prime\prime w}_{k+wj} & = & \delta^{\prime\prime}_{k+j} \rfloor_{h\rightarrow wh\mbox{,} x_{k+j}\rightarrow x_{k+wj}} \\  
\delta^{\prime\prime w}_{k+wj+l} & = & \delta^{\prime\prime w}_{k+wj}\rfloor_{\alpha(t)\rightarrow\alpha(t)-lh}\nonumber \\  
				 &   & \mbox{for } l \in [1,2, \ldots, w-1] \;.  
\end{eqnarray}  
These should be used with Eqs.\ (\ref{eq:gen_delta}),  
(\ref{eq:gen_delta_der1}) and (\ref{eq:gen_delta_der2}) to widen the  
Teukolsky source term by any factor $w$.  
  
\subsection{Smoothing the source with a Gaussian filter}  
\label{sec:smoothing}  
  
Further control of numerical noise can be achieved by filtering high  
frequency components in the source term.  This requires a convolution  
of the source with a discrete low pass filter.  We use a Gaussian  
filter because it maximizes the uncertainty principle --- it can be  
localized in both position and frequency with greatest efficiency.  
  
Consider a source of the form  
\begin{equation}  
s(x) = f_1(x)\delta(x-\alpha) + f_2(x)\delta^\prime(x-\alpha) +  
f_3(x)\delta^{\prime\prime}(x-\alpha)\;.  
\end{equation}  
Delta function identities allow us to rewrite this as  
\begin{equation}  
s(x) = g_1(\alpha)\delta(x-\alpha) + g_2(\alpha)\delta^\prime(x-\alpha) +  
g_3(\alpha)\delta^{\prime\prime}(x-\alpha)\;,  
\end{equation}  
where  
\begin{eqnarray}  
g_1(\alpha) &=& f_1(\alpha) - f_2'(\alpha) + f_3''(\alpha)\;,  
\nonumber\\  
g_2(\alpha) &=& f_2(\alpha) - 2f_3'(\alpha)\;,  
\nonumber\\  
g_3(\alpha) &=& f_3''(\alpha)\;.  
\end{eqnarray}  
On a discrete grid, this becomes  
\begin{equation}  
s(x_i) = s_i = g_1(\alpha)\delta_i + g_2(\alpha)\delta^\prime_i +  
g_3(\alpha)\delta^{\prime\prime}_i\;.    
\end{equation}  
If the delta function and its derivatives span $2n + 2$ grid points,  
with $x_{k + n - 1} \leq \alpha \leq x_{k + n}$, then $s_i \neq 0$ for  
$i \in [k - 1, \ldots, k + 2n]$. The source $s_i$ is zero everywhere  
else on the grid.  
  
The Gaussian filter is given by  
\begin{equation}  
c_k  =   
\frac{\exp[-\left(kh/b\right)^2/2]}  
{\sum_{i = -p}^{p}\exp[-\left(ih/b\right)^2/2]}\;,  
\end{equation}  
where $k \in [-p, -p+1, \dots, p]$ and $b$ is the width of the  
filter. The quantities $p$ and $b$ are adjustable parameters.  
Typically, we use $p=30$ and $b = 1.5h$.  Notice that  
\begin{equation}  
\sum_{i = -p}^{p} c_i = 1 \;;  
\end{equation}  
this normalization guarantees that the integrated value of any  
function convolved with the filter is unchanged.  
  
We now convolve the source with the filter to obtain  
\begin{eqnarray}  
sg_k & = & \sum_{i = -p}^{p} c_{i}s_{k + i} \;,  
\end{eqnarray}  
where $sg_k$ is the smoothed source term. This indicates that $sg_k  
\neq 0 $ for $k \in [k - p,\dots, k + 2n + p - 1]$.  
  
%
  
A wide filter spreads the source over a large domain on the numerical  
grid and thus increases errors, although it eliminates spurious  
harmonics.  We have found that using a wide stencil followed by a  
narrow Gaussian smoother works very well to reduce numerical noise and  
minimize errors from an insufficiently pointlike source.  
  
\subsection{Order of convergence of the filtered delta}  
\label{sec:Gaussconverge}  
  
Paper I discussed in detail the convergence of a code that uses a  
discrete delta.  Crucial background is given by Ref.\ {\cite{te04}}  
and summarized in Paper I.  The key point is that the moment  
\begin{equation}  
M_r = h\sum_{i = k}^{k + 2n - 1} \delta_i (x_i - \alpha)^r  
\end{equation}  
controls the delta's convergence properties.  Clearly, $M_0 = 1$  
(otherwise the delta is not properly normalized); in the continuum  
limit, $M_r = 0$ for $r > 0$.  For the discrete delta, the smallest  
non-zero value of $r$ for which $M_r \ne 0$ sets the order of  
convergence.  In particular, if $M_r \ne 0$, then a code which uses  
this delta will be no higher than $r$th-order convergent.  
  
We now show that, if a delta representation is second-order convergent  
before smoothing with the Gaussian filter ($M_0 = 1$, $M_1 = 0$, $M_2  
\ne 0$), it will remain second-order convergent after smoothing.  Upon  
convolving the discrete delta with the Gaussian smoother, we find  
\begin{eqnarray}  
\delta g_i & = & \sum_{j=-p}^{p} c_j \delta_{i+j}\;.  
\end{eqnarray}  
Let us denote the moments of the smoothed delta by $M^g_r$.  As  
discussed in Sec.\ {\ref{sec:smoothing}}, the convolution does not  
change the delta's normalization as long as the Gaussian filter is  
itself properly normalized; thus  
\begin{eqnarray}  
M^g_0 \equiv h \sum_{i=k}^{k+2n-1} \delta g_i & = & 1 \; .  
\end{eqnarray}  
We now examine the next higher moment of the smoothed delta:  
\begin{widetext}  
\begin{eqnarray}  
M^g_1 \equiv h \sum_{i=k-p}^{k+2n+p-1} \delta g_i (x_i-\alpha)  
& = & h \sum_{j=-p}^{p}   \sum_{i=k-p}^{k+2n+p-1} c_j\delta_{i+j}  
(x_i-\alpha) \;,\nonumber\\  
& = & h \sum_{j=-p}^{p} c_j \sum_{i=k-p}^{k+2n+p-1} \delta_{i+j}  
(x_i-\alpha)   \;,\nonumber\\  
& = & h \sum_{j=-p}^{p} c_j \sum_{i=k-p}^{k+2n+p-1} \delta_{i+j}  
(x_{i+j}-\alpha-jh)   \;,\nonumber\\  
& = &  h \sum_{j=-p}^{p} c_j \sum_{i=k-p}^{k+2n+p-1} \delta_{i+j}  
(x_{i+j}-\alpha) - h \sum_{j=-p}^{p} h j c_j  
\sum_{i=k-p}^{k+2n+p-1}\delta_{i+j} \;.  
\label{eq:first_mom}		  
\end{eqnarray}  
The first term on the final line of (\ref{eq:first_mom}) gives zero:  
Since $\sum\delta_l x_l = \alpha$,  
\begin{eqnarray}  
h \sum_{j=-p}^{p} c_j \sum_{i=k-p}^{k+2n+p-1} \delta_{i+j} (x_{i+j}-\alpha)  
&=& h \sum_{j=-p}^{p} c_j \sum_{l=k-p+j}^{k+2n+p+j-1}  
\delta_{l} (x_{l}-\alpha) \nonumber\\  
& = & 0 \;.  
\label{eq:first_mom_1}  
\end{eqnarray}  
The second line follows because $|j| <= p$, $\delta_i = 0$ if $i$ lies  
outside $[k,k+2n-1]$ and $M_1 = 0$.  
  
The second term on the final line of (\ref{eq:first_mom}) also yields  
zero:  
\begin{eqnarray}  
h \sum_{j = -p}^{p} h j c_j \sum_{i = k - p}^{k + 2n + p - 1}  
\delta_{i + j} &=& h^2 \sum_{j = -p}^{p} j c_j  
\sum_{l = k - p + j}^{k + 2n + p + j - 1}\delta_{l} \; ,\nonumber\\  
&=& h^2 \sum_{j=-p}^{p} j c_j \nonumber\\  
&=& 0 \; .  
\label{eq:first_mom_2}	  
\end{eqnarray}  
The Gaussian filter's symmetry property $c_j = c_{-j}$ has been  
applied in the last step.  Hence, we find $M_1^g = M_1 = 0$.  
\end{widetext}  
  
Evaluating the second moment proceeds similarly, but we find in the  
end terms involving $\sum_{j=-p}^{p} j^2 c_j$ which do not vanish.  
Thus, $M^g_2$ is the first non-vanishing moment of the discrete delta,  
demonstrating that the Gaussian-filtered discrete delta function  
exhibits second-order convergence.  The argument can be extended to  
the delta derivatives as well.  The smoothed Teukolsky source term  
will thus be second-order convergent.  
   
\section{Waveforms and comparisons for generic geodesic kerr orbits}  
\label{sec:comparisons}  
  
We now present the waveforms generated by a point particle in a  
geodesic orbit around a Kerr black hole.  The code used to generate  
these waves is discussed in detail in Paper I; the only important  
change to that discussion is that the source term uses the techniques  
presented in Sec.\ {\ref{sec:delta}} above.  We begin by reviewing  
Kerr black hole geodesics, sketching the numerical scheme used to  
solve the equations of motion.  We then examine different classes of  
eccentric and inclined orbits and compare the waveforms against those  
obtained from a frequency-domain code whose details are given in Ref.\  
\cite{dh06}.  We compute the correlation between the two waveforms in  
order to measure our level of agreement with frequency-domain  
waveforms.  
  
Our numerical grid is laid out in Boyer-Lindquist coordinats and uses  
$(\delta r,\delta\theta,\delta t) = (0.04M,\pi/60,0.02M)$ for the  
radial, angular and temporal resolutions.  The source term is  
constructed using Eqs.\ (\ref{eq:gen_delta}),  
(\ref{eq:gen_delta_der1}) and (\ref{eq:gen_delta_der2}) with  
$n_\theta$ in the range $3$--$9$ (depending on the orbit) for the  
angular delta-function and $n_r = 2$ for the radial delta.  We use a  
Gaussian filter of width $b = 1.5 \delta\theta$ to smooth higher  
harmonic noise.  
  
\subsection{Geodesics in Kerr spacetime}  
\label{sec:Kerr_geodesics}  
  
The source term for the time-domain code takes as input the worldline  
of the perturbation's source.  Here, we neglect radiation reaction and  
assume that the point particle follows a bound geodesic trajectory  
around the central massive black hole.  This bound trajectory can be  
computed by numerically integrating the geodesic equations.  We now  
briefly review how we massage the geodesic equations to put them into  
a form that makes for accurate numerical calculation; this material is  
presented in greater depth in Sec.\ IIC of Ref.\ {\cite{bfggh}}.  
  
The normal ``textbook'' presentation of the equations governing Kerr  
black hole geodesics is  
\begin{eqnarray}  
\Sigma^2\left(\frac{dr}{d\tau}\right)^2 &=& \left[E(r^2 + a^2) - a L_z  
\right]^2\nonumber\\  
& & - \Delta\left[r^2 + (L_z - a E)^2 + Q\right]  
\nonumber\\  
&\equiv& R(r)  
\label{eq:rdot}\\  
\Sigma^2\left(\frac{d\theta}{d\tau}\right)^2 &=& Q  
- \cos^2\theta\left[a^2(1 - E^2) + L_z^2/\sin^2\theta\right]  
\nonumber\\  
\label{eq:thetadot}\\  
\Sigma \frac{d\phi}{d\tau} &=& \frac{L_z}{\sin^2\theta} - a E  
+ \frac{a}{\Delta}\left[E(r^2 + a^2) - a L_z\right]  
\nonumber\\  
\label{eq:phidot}\\  
\Sigma \frac{dt}{d\tau} &=& a(L_z - aE\sin^2\theta)  
\nonumber\\  
& & + \frac{r^2 + a^2}{\Delta}\left[E(r^2 + a^2) - a L_z\right]\;.  
\label{eq:tdot}  
\end{eqnarray}  
[See, e.g., Ref.\ {\cite{mtw}}, Eqs.\ (33.32a--d).]  Here, $\Sigma =  
r^2 + a^2\cos^2\theta$, $\Delta = r^2 - 2 Mr + a^2$ (where $a = |\vec  
S|/M$ is the black hole's spin angular momentum per unit mass).  The  
constants of motion are orbital energy $E$, axial angular momentum  
$L_z$, and Carter constant $Q$.  
  
This form of the equations of motion is not well suited to numerical  
studies; in particular, $dr/d\tau$ and $d\theta/d\tau$ pass through  
zero and change sign when the orbiting body goes through turning  
points associated with those motions.  A handy way to eliminate these  
problems is to eliminate the turning points by remapping the  
coordinates $r$ and $\theta$ to parameters which accumulate secularly.  
The following parameterization, inspired by the Newtonian limit, has  
been found to work extremely well even deep in the strong field of  
rapidly rotating black holes:  
\begin{eqnarray}  
r &=& \frac{p}{1 + e\cos\psi}\;,  
\label{eq:psidef}\\  
\cos\theta &=& \cos\theta_{\min} \cos\chi\;.  
\label{eq:chidef}  
\end{eqnarray}  
In the Newtonian limit, $p$ is the orbit's semi-latus rectum, and $e$  
is its eccentricity; $\theta_{\min}$ is the minimum value of $\theta$  
reached by the orbiting body, and is used to define the orbit's  
inclination $\theta_{\rm inc}$  
\begin{equation}  
\theta_{\rm inc} = \frac{\pi}{2} - {\rm sgn}(L_z) \theta_{\min}\;.  
\end{equation}  
Once $E$, $L_z$, and $Q$ are specified, $p$, $e$, and $\theta_{\rm  
inc}$ are fully determined.  It is then a straightforward matter to  
turn Eqs.\ (\ref{eq:rdot}) and (\ref{eq:thetadot}) into expressions  
for $d\psi/d\tau$ and $d\chi/d\tau$; see Ref.\ {\cite{bfggh}} for  
details.  The resulting expressions behave extremely well for all  
bound orbits outside the black hole's event horizon.  A numerical  
integrator for these variables allows us to compute the dynamics of  
our orbiting body's Teukolsky equation source term.  
  
Before moving on, we note that, within the context of the  
dissipative-only or radiative approximation to inspiral, it is simple  
to modify these equations to build the worldline of an inspiralling  
body: We simply allow the orbital ``constants'' ($E$, $L_z$, and $Q$;  
or, $p$, $e$, and $\theta_{\rm inc}$) to evolve according to the  
inspiral law.  Reference {\cite{bfggh}} uses approximate radiation  
reaction, based on fits to strong-field radiation reaction  
calculations in regimes where it is well understood, to compute the  
inspiral worldlines which underlie the ``kludge'' waveforms.  We use  
this prescription for evolving the constants in Sec.\  
{\ref{sec:inspiral}} to demonstrate this code's ability to compute  
inspiral waves.  
  
\subsection{Comparison with frequency-domain waveforms}  
\label{sec:comparison}  
  
To validate our waveforms, we compare with the ``snapshots'' generated  
using the frequency-domain code described in Ref.\ {\cite{dh06}}.  
This code uses the fact that bound Kerr geodesics are fully described  
by three frequencies (radial $\Omega_r$, latitudinal $\Omega_\theta$,  
and axial $\Omega_\phi$) to build the waveform from a geodesic orbit  
as a sum over harmonics of these frequencies {\cite{dh04}}.  Since  
both the time-domain and frequency-domain codes solve the same master  
equation, they should produce identical waveforms for identical  
orbits, so long as each code is sufficiently accurate.  
  
To quantify the accuracy with which a time-domain waveform $X = (x_1,  
x_2, \ldots, x_n)$ agrees with a frequency-domain waveform $Y = (y_1,  
y_2, \ldots, y_n)$, we use the following correlation measure:  
\begin{eqnarray}  
\label{eq:corr}  
r_{XY} \equiv \frac{\sum(x_i - \bar x)(y_i - \bar y)}  
{\sqrt{\sum(x_i - \bar x)^2}\sqrt{\sum(y_i - \bar y)^2}}\;.  
\end{eqnarray}  
(The sums in all cases are from $i = 1$ to $i = n$.)  This coefficient  
is identical to the {\it match} between two waveforms defined by Owen  
{\cite{owen96}} in the white noise limit [noise spectral density  
$S_h(f) = {\rm constant}$].  One might expect the waveforms' mean  
values $\bar x$ and $\bar y$ to equal zero.  However, finite duration  
effects can make these quantities slightly non-zero, so it is useful  
to explicitly do this subtraction.  
  
A useful reformulation of Eq.\ (\ref{eq:corr}) is  
\begin{eqnarray}  
r_{XY} \equiv \frac{n\sum x_i y_i - \sum x_i \sum y_i}{\sqrt{n\sum  
x_i^2 - \left(\sum x_i\right)^2}\sqrt{n\sum y_i^2 - \left(\sum  
y_i\right)^2 }}\;.  
\end{eqnarray}  
Note that $r_{XY}$ is always between $-1$ and $1$; a value close to  
$1$ indicates that the two waveforms are well correlated.  Note also  
that the correlation depends on how many points $n$ are used in  
comparing the two waveforms (or equivalently, the span of time over  
which we compare the waves).  We have found that as long as $n  
\gtrsim$ several hundred, we get consistent results: Changing $n$ for  
a given comparison only causes small variations in the fourth  
significant digit of $r_{XY}$.  
  
It is of course possible to concoct other measures of how well two  
waveforms agree.  Ideally, disagreements between waveforms should be  
quantified in terms of their observational significance.  For example,  
Cutler and Vallisneri have demonstrated that it is not unusual for  
waveforms with a match of 0.9999 to differ significantly in their  
estimates of the parameters which describe the source {\cite{cutval}}.  
For our present purpose, $r_{XY}$ is sufficient to demonstrate that  
our time-domain code produces high quality waveforms; whether they are  
sufficiently high quality to be used for GW measurement purposes will  
need to be re-examined at a later time.  
  
An important step in producing accurate waveforms is to perform runs  
at multiple resolutions, then estimate (and eliminate) the waveform  
error using a form of Richardson extrapolation {\cite{numrec}}.  This  
plays a crucial role in reducing ``noise'' from spurious excitation of  
the large black hole's quasinormal modes.  The details of this  
extrapolation technique are described in Appendix\ \ref{sec:app}.  
  
\begin{figure*}[htb]  
\begin{center}  
\includegraphics[height = 100mm]{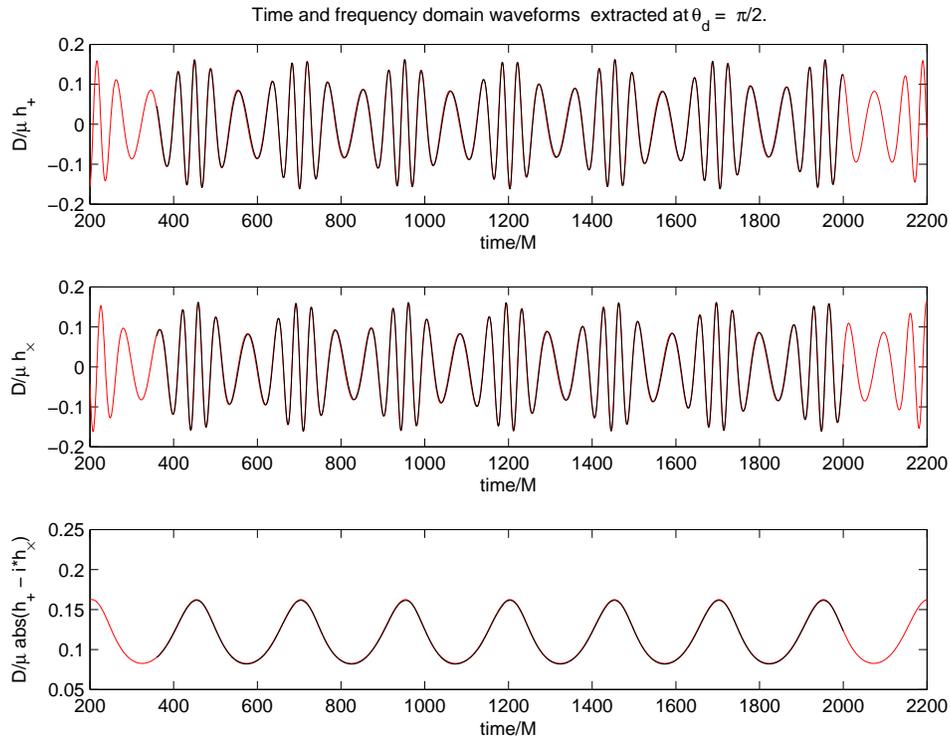}  
\caption{\label{tdfd_m2_eqecc}Comparison of time- and frequency-domain  
waveforms.  We show waves for the $m = 2$ mode from a point particle  
with orbital parameters $p = 6.472M$, $e = 0.3$ and $\theta_{\rm inc}  
= 0$ orbiting a black hole with spin $a/M = 0.3$.  The angle between  
the spin axis of the black hole and the line of sight is $\theta_d =  
\pi/2$.  Time-domain results are in black, frequency-domain results in  
red.  Top panel: ``plus'' polarizations in dimensionless units.  
Middle: ``cross'' polarizations. Bottom: Comparison of $|h_+ -  
ih_\times|$.  This last quantity gives a good visual measure of the  
level of agreement between the two waveforms.  The correlations  
between the two waveforms are $0.9974$ (plus) and $0.9975$ (cross).}  
\end{center}  
\end{figure*}  
  
\begin{figure*}[htb]  
\begin{center}  
\includegraphics[height = 100mm]{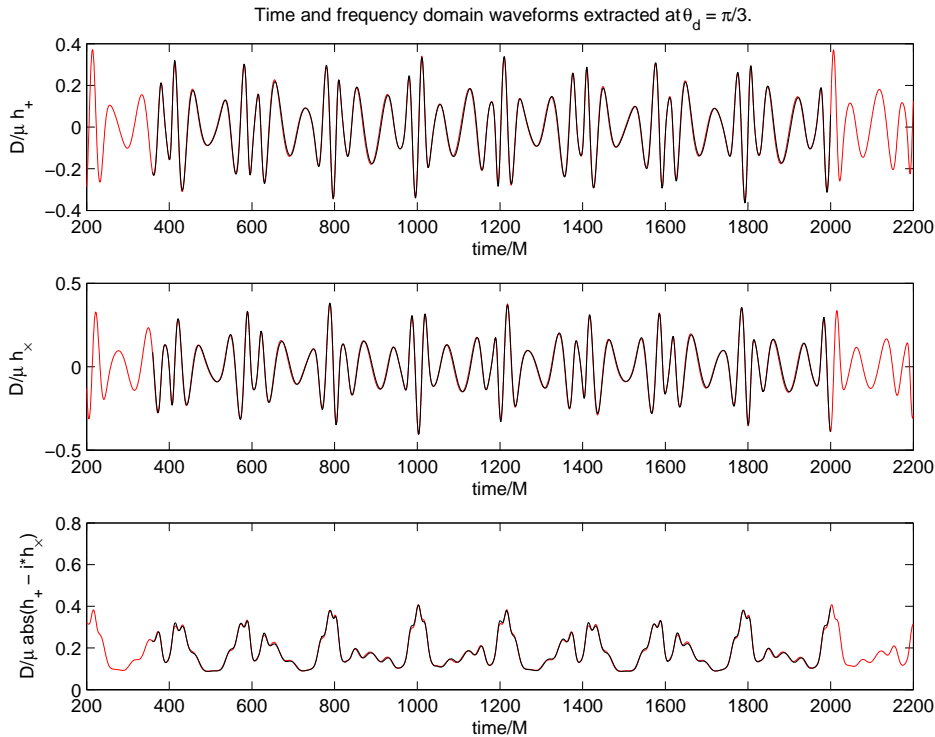}  
\caption{\label{tdfd_m2_gen}Comparison of time- and frequency-domain  
waveforms.  Here, we show waves for the $m = 2$ mode for a geodesic  
with $p = 6M$, $e = 0.3$ and $\theta_{\rm inc} = \pi/3$ about a black  
hole with spin $a/M = 0.9$; black is time-domain results, red is  
frequency domain.  The correlations in this case are $0.9961$ (plus)  
and $0.9962$ (cross).}  
\end{center}  
\end{figure*}  
  
\begin{figure*}[htb]  
\begin{center}  
\includegraphics[height = 100mm]{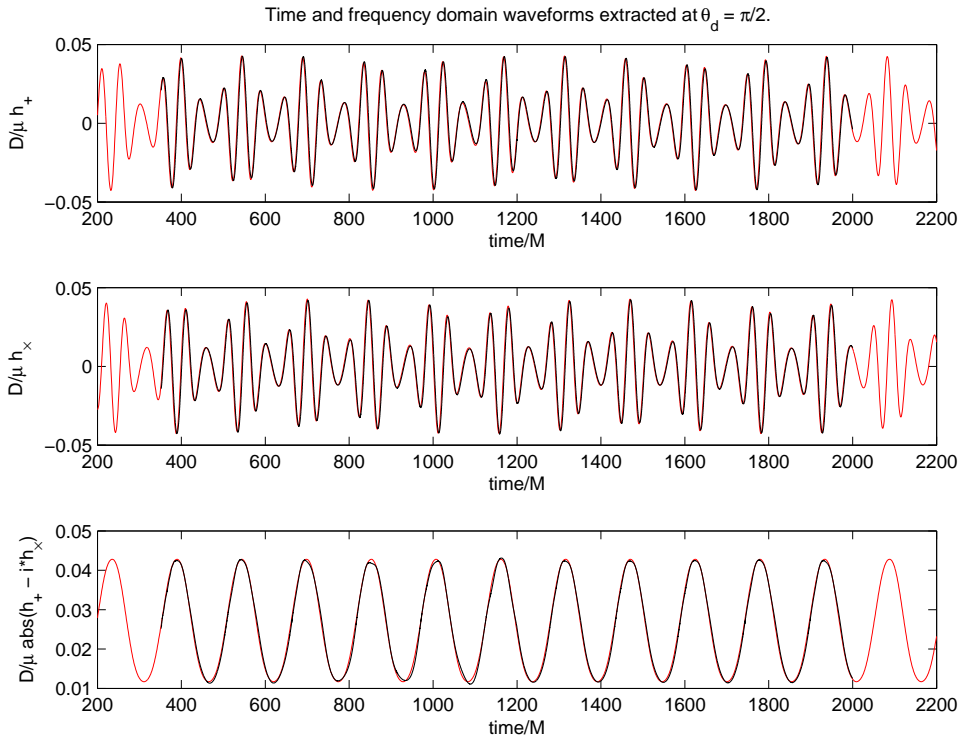}  
\caption{\label{tdfd_m3_incl}Comparison of time- and frequency-domain  
waveforms.  These waves are for the $m = 3$ mode from a circular  
geodesic with orbital parameters $p = 6M$, and $\theta_{\rm inc} =  
\pi/4$ around a hole with spin $a/M = 0.9$. All symbols have the same  
meaning as in Fig.\ \ref{tdfd_m2_eqecc}.  The correlations are  
$0.9769$ (plus) and $0.9770$ (cross).}  
\end{center}  
\end{figure*}  
  
\begin{table}[htb]
\caption{\label{corr_coeff_eq_m2} Correlation between time- and  
frequency-domain waveforms for the $m = 2$ mode for a range of  
equatorial, eccentric orbits.  The parameters $p$, $e$ and  
$\theta_{\rm inc}$ are semi-latus rectum, eccentricity, and  
inclination of the geodesic orbit, $a/M$ is the black hole spin and  
$\theta_d$ is the angle between the spin axis and the line of sight to  
the observer.  The last two columns show correlations for the plus and  
cross polarizations.}  
\begin{tabular}{|c|c|c|c|c|c|c|}  
\hline  
\hline  
$p/M$ & $e$ & $\theta_{\rm inc}$ (deg) & $a/M$ & $\theta_d$ (deg) & $h_+$ corr. & $h_\times$ corr. \\  
\hline  
$6.472$ & $0.3$ & $0$ & $0.3$ & $30$ & $0.9961$ & $0.9962$\\  
$6.472$ & $0.3$ & $0$ & $0.3$ & $60$ & $0.9969$ & $0.9969$\\  
$6.472$ & $0.3$ & $0$ & $0.3$ & $90$ & $0.9974$ & $0.9975$\\  
\hline  
$5.768$ & $0.3$ & $0$ & $0.7$ & $30$ & $0.9971$ & $0.9971$\\  
$5.768$ & $0.3$ & $0$ & $0.7$ & $60$ & $0.9977$ & $0.9978$\\  
$5.768$ & $0.3$ & $0$ & $0.7$ & $90$ & $0.9983$ & $0.9983$\\  
\hline  
$6.472$ & $0.7$ & $0$ & $0.3$ & $30$ & $0.9915$ & $0.9911$ \\  
$6.472$ & $0.7$ & $0$ & $0.3$ & $60$ & $0.9911$ & $0.9908$\\  
$6.472$ & $0.7$ & $0$ & $0.3$ & $90$ & $0.9900$ & $0.9901$\\  
\hline  
$5.768$ & $0.7$ & $0$ & $0.7$ & $30$ & $0.9625$ & $0.9607$ \\  
$5.768$ & $0.7$ & $0$ & $0.7$ & $60$ & $0.9621$ & $0.9601$ \\  
$5.768$ & $0.7$ & $0$ & $0.7$ & $90$ & $0.9596$ & $0.9578$ \\  
\hline  
\hline  
\end{tabular}  
\end{table}  
  
\begin{table}[ht]
\caption{\label{corr_coeff_incl_m2}Correlation between time- and  
frequency-domain waveforms for the $m = 2$ mode for a range of  
inclined nearly circular orbits. All symbols have the same meaning as  
in Table \ref{corr_coeff_eq_m2}.}  
\begin{tabular}{|c|c|c|c|c|c|c|}  
\hline  
\hline  
$p/M$ & $e$ & $\theta_{\rm inc}$ (deg) & $a/M$ & $\theta_d$ (deg) & $h_+$ corr. & $h_\times$ corr. \\  
\hline  
$6$ & $10^{-4}$ & $45$ & $0.5$ & $60$ & $0.9968$ & $0.9967$ \\  
$6$ & $10^{-4}$ & $45$ & $0.5$ & $90$ & $0.9961$ & $0.9960$ \\  
$8$ & $10^{-4}$ & $45$ & $0.5$ & $60$ & $0.9923$ & $0.9919$ \\  
$8$ & $10^{-4}$ & $45$ & $0.5$ & $90$ & $0.9908$ & $0.9903$ \\  
\hline  
$6$ & $10^{-4}$ & $45$ & $0.9$ & $60$ & $0.9967$ & $0.9967$ \\  
$6$ & $10^{-4}$ & $45$ & $0.9$ & $90$ & $0.9961$ & $0.9961$ \\  
$8$ & $10^{-4}$ & $45$ & $0.9$ & $60$ & $0.9920$ & $0.9919$ \\  
$8$ & $10^{-4}$ & $45$ & $0.9$ & $90$ & $0.9905$ & $0.9907$ \\  
\hline  
$6$ & $10^{-4}$ & $60$ & $0.5$ & $60$ & $0.9964$ & $0.9965$ \\  
$6$ & $10^{-4}$ & $60$ & $0.5$ & $90$ & $0.9952$ & $0.9952$ \\  
$8$ & $10^{-4}$ & $60$ & $0.5$ & $60$ & $0.9917$ & $0.9910$ \\  
$8$ & $10^{-4}$ & $60$ & $0.5$ & $90$ & $0.9888$ & $0.9882$ \\  
\hline  
$6$ & $10^{-4}$ & $60$ & $0.9$ & $60$ & $0.9986$ & $0.9986$ \\  
$6$ & $10^{-4}$ & $60$ & $0.9$ & $90$ & $0.9981$ & $0.9982$ \\  
$8$ & $10^{-4}$ & $60$ & $0.9$ & $60$ & $0.9917$ & $0.9915$ \\  
$8$ & $10^{-4}$ & $60$ & $0.9$ & $90$ & $0.9891$ & $0.9890$ \\  
\hline  
\hline  
\end{tabular}  
\end{table}  
  
\begin{table}[ht]
\caption{\label{corr_coeff_gen_m2} Correlation between time- and  
frequency-domain waveforms for the $m = 2$ mode for a range of generic  
orbits. All symbols have the same meaning as in Table  
\ref{corr_coeff_eq_m2}.}  
\begin{tabular}{|c|c|c|c|c|c|c|}  
\hline  
\hline  
$p/M$ & $e$ & $\theta_{\rm inc}$ (deg) & $a/M$ & $\theta_d$ (deg) & $h_+$ corr. & $h_\times$ corr. \\  
\hline  
$6$ & $0.3$ & $40$ & $0.9$ & $60$ & $0.9978$ & $0.9978$ \\  
$6$ & $0.3$ & $40$ & $0.9$ & $90$ & $0.9976$ & $0.9976$ \\  
$8$ & $0.3$ & $40$ & $0.5$ & $60$ & $0.9898$ & $0.9897$ \\  
$8$ & $0.3$ & $40$ & $0.5$ & $90$ & $0.9910$ & $0.9910$ \\  
\hline  
$6$ & $0.7$ & $40$ & $0.9$ & $60$ & $0.9898$ & $0.9906$ \\  
$6$ & $0.7$ & $40$ & $0.9$ & $90$ & $0.9889$ & $0.9891$ \\  
$6$ & $0.7$ & $60$ & $0.9$ & $60$ & $0.9905$ & $0.9868$ \\  
$6$ & $0.7$ & $60$ & $0.9$ & $90$ & $0.9895$ & $0.9866$ \\  
\hline  
$6$ & $0.3$ & $60$ & $0.9$ & $60$ & $0.9961$ & $0.9962$ \\  
$6$ & $0.3$ & $60$ & $0.9$ & $90$ & $0.9950$ & $0.9954$ \\  
$8$ & $0.3$ & $60$ & $0.5$ & $60$ & $0.9906$ & $0.9890$ \\  
$8$ & $0.3$ & $60$ & $0.5$ & $90$ & $0.9884$ & $0.9866$ \\  
\hline  
\hline  
\end{tabular}  
\end{table}  
  
\begin{table}[ht]
\caption{\label{corr_coeff_eq_m3} Correlation between time- and  
frequency-domain waveforms for the $m = 3$ mode for a range of  
equatorial eccentric orbits.  All symbols are as in Table  
\ref{corr_coeff_eq_m2}.}  
\begin{tabular}{|c|c|c|c|c|c|c|}  
\hline  
\hline  
$p/M$ & $e$ & $\theta_{\rm inc}$ (deg) & $a/M$ & $\theta_d$ (deg) & $h_+$ corr. & $h_\times$ corr. \\  
\hline  
$6.472$ & $0.3$ & $0$ & $0.3$ & $30$ & $0.9908$ & $0.9909$ \\  
$6.472$ & $0.3$ & $0$ & $0.3$ & $60$ & $0.9922$ & $0.9922$ \\  
$6.472$ & $0.3$ & $0$ & $0.3$ & $90$ & $0.9930$ & $0.9931$ \\  
\hline  
$5.768$ & $0.3$ & $0$ & $0.7$ & $30$ & $0.9934$ & $0.9935$ \\  
$5.768$ & $0.3$ & $0$ & $0.7$ & $60$ & $0.9943$ & $0.9944$ \\  
$5.768$ & $0.3$ & $0$ & $0.7$ & $90$ & $0.9948$ & $0.9948$ \\  
\hline  
$6.472$ & $0.7$ & $0$ & $0.3$ & $30$ & $0.9931$ & $0.9931$ \\  
$6.472$ & $0.7$ & $0$ & $0.3$ & $60$ & $0.9905$ & $0.9906$ \\  
$6.472$ & $0.7$ & $0$ & $0.3$ & $90$ & $0.9923$ & $0.9923$ \\  
\hline  
$5.768$ & $0.7$ & $0$ & $0.7$ & $30$ & $0.9928$ & $0.9929$\\  
$5.768$ & $0.7$ & $0$ & $0.7$ & $60$ & $0.9932$ & $0.9930$\\  
$5.768$ & $0.7$ & $0$ & $0.7$ & $90$ & $0.9920$ & $0.9921$\\  
\hline  
\hline  
\end{tabular}  
\end{table}  
  
\begin{table}[ht]
\caption{\label{corr_coeff_incl_m3} Correlation between time- and  
frequency-domain waveforms for the $m = 3$ mode for a range of  
inclined nearly circular orbits. All symbols are as in Table  
\ref{corr_coeff_eq_m2}.}  
\begin{tabular}{|c|c|c|c|c|c|c|}  
\hline  
\hline  
$p/M$ & $e$ & $\theta_{\rm inc}$ (deg) & $a/M$ & $\theta_d$ (deg) & $h_+$ corr. & $h_\times$ corr. \\  
\hline  
$6$ & $10^{-4}$ & $45$ & $0.5$ & $60$ & $0.9918$ & $0.9918$ \\  
$6$ & $10^{-4}$ & $45$ & $0.5$ & $90$ & $0.9907$ & $0.9907$ \\  
$8$ & $10^{-4}$ & $45$ & $0.5$ & $60$ & $0.9798$ & $0.9798$ \\  
$8$ & $10^{-4}$ & $45$ & $0.5$ & $90$ & $0.9773$ & $0.9772$ \\  
\hline  
$6$ & $10^{-4}$ & $45$ & $0.9$ & $60$ & $0.9912$ & $0.9913$ \\  
$6$ & $10^{-4}$ & $45$ & $0.9$ & $90$ & $0.9905$ & $0.9906$ \\  
$8$ & $10^{-4}$ & $45$ & $0.9$ & $60$ & $0.9787$ & $0.9790$ \\  
$8$ & $10^{-4}$ & $45$ & $0.9$ & $90$ & $0.9769$ & $0.9770$ \\  
\hline  
$6$ & $10^{-4}$ & $60$ & $0.5$ & $60$ & $0.9884$ & $0.9884$ \\  
$6$ & $10^{-4}$ & $60$ & $0.5$ & $90$ & $0.9876$ & $0.9876$ \\  
$8$ & $10^{-4}$ & $60$ & $0.5$ & $60$ & $0.9636$ & $0.9640$ \\  
$8$ & $10^{-4}$ & $60$ & $0.5$ & $90$ & $0.9674$ & $0.9675$\\  
\hline  
$6$ & $10^{-4}$ & $60$ & $0.9$ & $60$ & $0.9665$ & $0.9661$ \\  
$6$ & $10^{-4}$ & $60$ & $0.9$ & $90$ & $0.9680$ & $0.9678$ \\  
$8$ & $10^{-4}$ & $60$ & $0.9$ & $60$ & $0.9463$ & $0.9473$ \\  
$8$ & $10^{-4}$ & $60$ & $0.9$ & $90$ & $0.9608$ & $0.9641$ \\  
\hline  
\hline  
\end{tabular}  
\end{table}  
  
\begin{table}[ht]
\caption{\label{corr_coeff_gen_m3} Correlation between time- and  
frequency-domain waveforms for the $m = 3$ mode for a range of generic  
orbits. All symbols are as in Table \ref{corr_coeff_eq_m2}.}  
\begin{tabular}{|c|c|c|c|c|c|c|}  
\hline  
\hline  
$p/M$ & $e$ & $\theta_{\rm inc}$ (deg) & $a/M$ & $\theta_d$ (deg) & $h_+$ corr. & $h_\times$ corr. \\  
\hline  
$6$ & $0.3$ & $40$ & $0.9$ & $60$ & $0.9917$ & $0.9916$ \\  
$6$ & $0.3$ & $40$ & $0.9$ & $90$ & $0.9915$ & $0.9914$ \\  
$8$ & $0.3$ & $40$ & $0.5$ & $60$ & $0.9801$ & $0.9803$ \\  
$8$ & $0.3$ & $40$ & $0.5$ & $90$ & $0.9785$ & $0.9785$ \\  
\hline  
$6$ & $0.7$ & $40$ & $0.9$ & $60$ & $0.9906$ & $0.9981$ \\  
$6$ & $0.7$ & $40$ & $0.9$ & $90$ & $0.9899$ & $0.9895$ \\  
$6$ & $0.7$ & $60$ & $0.9$ & $60$ & $0.9862$ & $0.9862$ \\  
$6$ & $0.7$ & $60$ & $0.9$ & $90$ & $0.9819$ & $0.9821$ \\  
\hline  
$6$ & $0.3$ & $60$ & $0.9$ & $60$ & $0.9790$ & $0.9788$ \\  
$6$ & $0.3$ & $60$ & $0.9$ & $90$ & $0.9840$ & $0.9839$ \\  
$8$ & $0.3$ & $60$ & $0.5$ & $60$ & $0.9788$ & $0.9791$ \\  
$8$ & $0.3$ & $60$ & $0.5$ & $90$ & $0.9747$ & $0.9744$ \\  
\hline  
\hline  
\end{tabular}  
\end{table}  
  
Tables \ref{corr_coeff_eq_m2}, \ref{corr_coeff_incl_m2},  
\ref{corr_coeff_gen_m2}, \ref{corr_coeff_eq_m3},  
\ref{corr_coeff_incl_m3} and \ref{corr_coeff_gen_m3} list the  
correlation coefficients for the $m=2$ and $m=3$ azimuthal modes of  
different classes of orbits.  The coefficient is greater than $0.99$  
for a large fraction of parameter space. Time domain runs corresponding to each column required about 125 CPU hours on an Apple MacPro processor.  That code was compiled using the Intel C++
compiler.  The frequency domain code's cost is about 3-4 CPU hours per
waveform when attempting to get both asymptotic energy fluxes to accuracies
of about $0.1\%$ to $1\%$ on a machine using a 3.2 GHz Pentium 4 Xeon processor. We also show (Figs.\  
\ref{tdfd_m2_eqecc}, \ref{tdfd_m2_gen}, and \ref{tdfd_m3_incl})  
examples of the waves, computed with both time- and frequency-domain  
codes, to give the reader a visual sense of the overlap.  
  
\section{Inspiral waveforms}  
\label{sec:inspiral}  
  
Having demonstrated that the finite-impulse source works well for  
astrophysically relevant generic black hole orbits, we now examine how  
well we do evolving through a sequence of such orbits.  Since each  
orbit in the sequence is no different than the orbits that we  
validated against in Sec.\ {\ref{sec:comparison}}, we anticipate no  
great difficulty here.  Indeed, the biggest challenge is choosing a  
method to evolve through our sequence.  Our goal is to do this with a  
frequency-domain code to build the orbital-constant trajectory  
$[E(t),L_z(t),Q(t)]$.  To quickly produce results that are  
qualitatively correct, we presently make this trajectory using the  
``kludge'' inspiral treatment described in Ref.\ {\cite{gg06}}, and  
used to make model waveforms in Ref.\ {\cite{bfggh}}.  The ``kludge''  
uses a somewhat idiosyncratic mix of post-Newtonian backreaction  
formulae combined with numerical results from frequency-domain  
backreaction in the circular, inclined ($e = 0$, $\theta_{\rm inc} \ne  
0$) and eccentric, equatorial ($e \ne 0$, $\theta_{\rm inc} = 0$)  
limits to estimate the properties of EMRI waves.  By construction, the  
results agree very well with Teukolsky-based inspirals in those  
limits; for the generic case, they produce plausible inspirals.  
  
\begin{figure*}[htb]  
\begin{center}  
\includegraphics[height = 130mm]{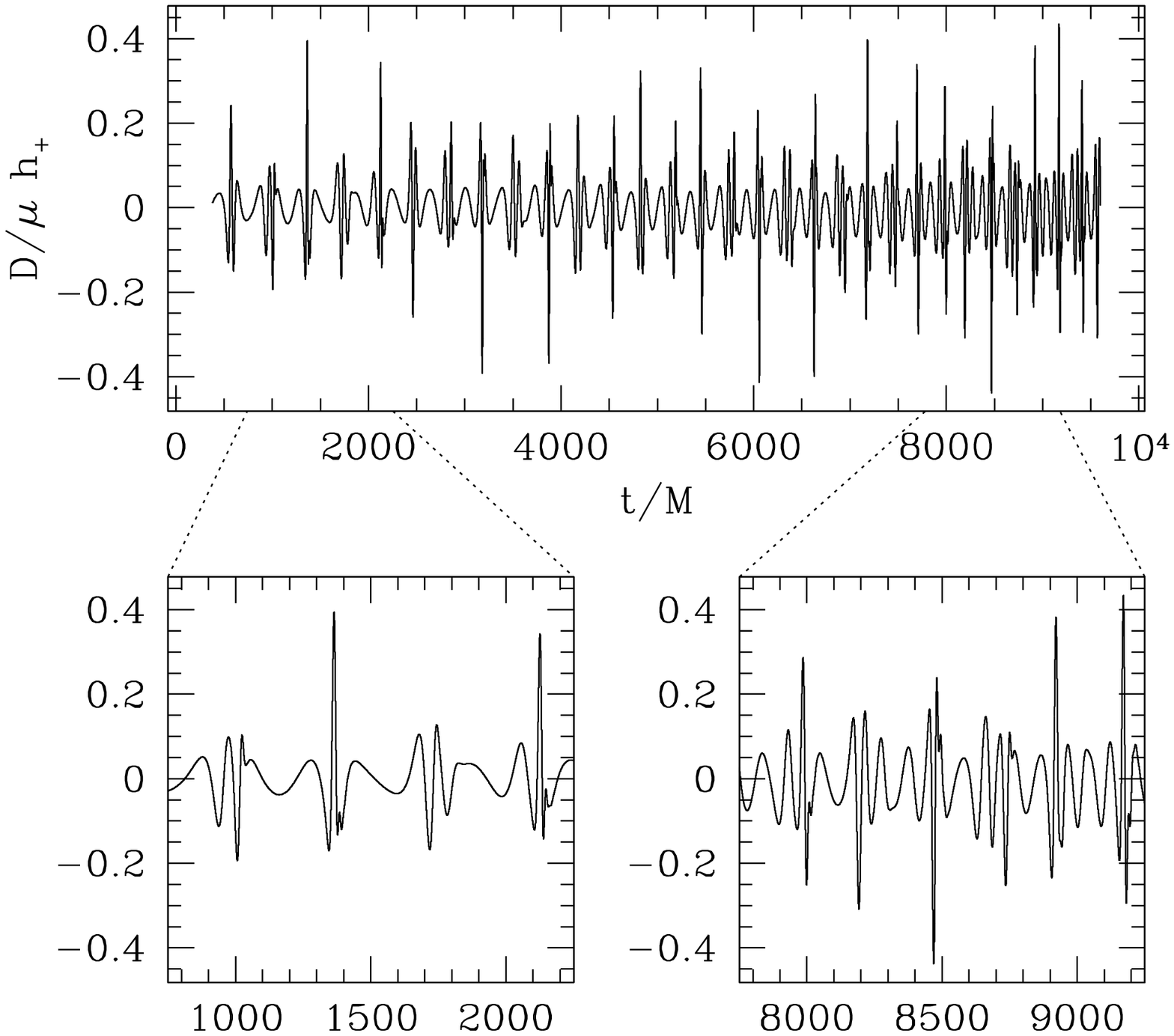}  
\vskip -1cm  
\caption{\label{inspiral_wave} Waveform ($m = 2$ mode) of a small body  
spiraling into a massive black hole.  We use ``kludge'' backreaction  
to evolve through a sequence of orbits, but compute the waves with our  
time-domain solver.  The large black hole has spin $a = 0.5M$; the  
small body's orbit initially has parameters $p = 10M$, $e = 0.5$, and  
$\theta_{\rm inc} = 0.5$ radians.  The mass ratio of the system is  
$\mu/M = 0.016$.  The top panel shows the full span that we simulated;  
the bottom two panels are zooms on early (bottom left) and late  
(bottom right) segments.  Note the clear evolution of the wave's  
frequency as the orbit's mean radius shrinks.}  
\end{center}  
\end{figure*}  
  
Figure {\ref{inspiral_wave}} shows our waveform for a ``kludge''  
inspiral.  We took the large black hole to have spin $a = 0.5M$, and  
set the mass ratio to $\mu/M = 0.016$.  The orbit was initially chosen  
to have semi-latus rectum $p = 10M$, eccentricity $e = 0.5$, and  
inclination $\theta_{\rm inc} = 0.5$ radians.  This figure shows  
features reminiscent of the geodesic snapshots shown in Figs.\  
{\ref{tdfd_m2_eqecc}}, {\ref{tdfd_m2_gen}}, and {\ref{tdfd_m3_incl}};  
in addition, one can clearly see evolution of the wave's properties.  
The increase in the wave's frequency, largely due to the decay of the  
orbit's semi-latus rectum, is quite clear.  Perhaps less obvious is a  
signature of the eccentricity's decay.  This is illustrated most  
clearly by comparing the lower left and lower right panels of Fig.\  
{\ref{inspiral_wave}}, which zoom onto early and late portions of the  
inspiral.  Early on, the waveform is dominated by a series of  
high-frequency bursts; these occur when the small body passes through  
periapsis and ``whirls'' most rapidly about the massive black hole.  
There is then a relatively quiet section while the body ``zooms'' out  
to apoapsis, and then comes in to ``whirl'' at periapsis again.  As  
eccentricity shrinks, the difference between periapsis and apoapsis  
becomes smaller.  The high-frequency bursts crowd closer and closer  
together, approaching a continuum sinusoid as the eccentricity  
approaches zero.  
  
Although this inspiral model is somewhat unphysical, we expect that it  
shares many properties with true adiabatic inspiral waveforms.  In  
particular, the spectral evolution of a wave like that in Fig.\  
{\ref{inspiral_wave}} should be quite similar to the evolution of real  
EMRI waveforms.  It should be emphasized that computing the waveform  
shown in Fig.\ {\ref{inspiral_wave}} required about as much  
computational effort as computing the geodesic snapshot waves, Figs.\  
{\ref{tdfd_m2_eqecc}}, {\ref{tdfd_m2_gen}}, and {\ref{tdfd_m3_incl}}  
(modulo a factor $\sim 4$--$5$ since the waveform in Fig.\  
{\ref{inspiral_wave}} lasts $\sim 4$--$5$ times longer than the  
others).  Given a robust code to generate the inspiral worldline of  
EMRI systems, the waveforms that our code produces should be a useful  
tool for examining issues in LISA measurement and data analysis.  
  
\section{Summary and future work}  
\label{sec:summary}  
  
We have now shown that the finite impulse delta representation of the  
time-domain Teukolsky equation's source works very well for  
complicated and astrophysically relevant orbits.  In our previous  
analysis (\cite{skh07}, Paper I), we confined ourselves to the  
simplest circular, equatorial black hole orbits.  The basic ideas from  
Paper I work well even when the source arises from highly inclined and  
highly eccentric orbits, and when the source evolves through a  
sequence of those orbits.  It is now a relatively straightforward  
matter to compute the waves arising from a body following any  
reasonably behaved worldline in the spacetime of a black hole.  
  
The primary complication arising from these more generic orbit classes  
is that the orbiting body will cross zones within our numerical grid.  
The source thus becomes dynamical; the finite-impulse delta must  
likewise be dynamical to represent it.  The evolution of the impulses  
that we use to represent the delta can seed numerical noise, reducing  
the calculation's accuracy.  We have found that minor extensions of  
Paper I's basic techniques greatly mitigate the impact of this source  
of numerical noise.  In particular, by using a higher-order  
representation (Sec.\ \ref{sec:highorder}), the delta is smoothed  
enough that the coupling to the Teukolsky equation's second-order  
differential operators does not seed much error.  Widening the delta's  
stencil (Sec.\ \ref{sec:widestencils}) also helps, since the  
fractional change in a given impulse will be less if the delta is  
represented by more impulses.  Finally, residual high frequency noise  
not removed by these techniques can be taken out by convolving the  
Teukolsky source term with a low-pass (Gaussian) filter (Sec.\  
\ref{sec:smoothing}).  It's worth emphasizing that we smooth the  
entire source term, not just the delta function (which would arguably  
make our delta rather similar to the truncated Gaussian  
{\cite{k04,bk07}} which this technique was designed to improve upon).  
  
Comparison with results from the frequency-domain {\cite{dh06}}  
demonstrates that the waveforms generated with this source term are of  
very high quality (Sec.\ \ref{sec:comparisons}).  Visually, the  
waveforms lie on top of one another in every case that we have  
examined; a quantitative overlap integral demonstrates that waveforms  
from the two calculations are often more than $99\%$ correlated.  A  
key step in achieving such high quality results is to estimate the  
largest errors in our time-domain calculations, and then subtract that  
estimate from our result.  We do this by performing these calculations  
at two different grid resolutions; under the assumption that our  
dominant error is quadratic in grid spacing, we then estimate the  
magnitude of our error (Appendix {\ref{sec:app}}).  The excellent  
agreement we achieve with frequency-domain results validates this  
approach, at least for all the cases we have considered.  
  
So far, our main physics accomplishment is excellent agreement between  
time- and frequency-domain approaches to waveform calculation.  It  
should be emphasized, however, that for waveform calculations, there  
will be a large set of circumstances in which time-domain codes are  
more efficient.  For generic orbits, a frequency-domain code may  
require the calculation and summation of many thousand multipoles and  
Fourier modes.  A time-domain code ``automatically'' sums over all  
modes (except the $m$ index), so that (in principle) it is no more  
difficult to compute the waves from a highly inclined, highly  
eccentric black hole orbit than from an orbit with modest inclination  
and eccentricity.  
  
The real payoff of this tool will come when we allow the source to  
radiatively decay, evolving through a sequence of orbits.  As a  
demonstration that this can be done, we use a ``kludged'' inspiral to  
compute a body's inspiral, and then use that inspiral as the source  
for our time-domain solver in Sec.\ {\ref{sec:inspiral}}.  Though not  
a physically accurate inspiral, this scenario shares many properties  
with the actual adiabatic inspiral.  In particular, it demonstrates  
the computational advantage of a robust time-domain code for computing  
inspiral waveforms, given the worldline the inspiraling body follows.  
  
Future work will address our goal of complete waveforms for the EMRI  
problem, in the context of the dissipation-only approximation to EMRI  
dynamics.  We have recently extended our frequency-domain code to  
include the evolution of Carter's constant in the radiative  
backreaction limit {\cite{sthgn06}}, and will use this code to produce  
the radiation reaction data describing an inspiraling body.  With this  
step in hand, no issue of principle stands in the way of coupling the  
time- and frequency-domain approaches to make usefully accurate EMRI  
waveforms.  
  
\acknowledgments  
  
We are very grateful to Jonathan Gair and Kostas Glampedakis for their  
permission to use the code from Ref.\ {\cite{gg06}} to build the  
inspiral we use in Sec.\ {\ref{sec:inspiral}}.  P.~A.~S.\ and  
S.~A.~H.\ are supported by NASA Grant No.\ NNG05G105G; S.~A.~H.\ is  
additionally supported by NSF Grant No.\ PHY-0449884 and the MIT Class  
of 1956 Career Development fund.  G.~K.\ acknowledges research support  
from the University of Massachusetts and the Fund for Astrophysical Research, Inc.,  
as well as supercomputing support from the TeraGrid (Grant No.\ TG-PHY060047T), which was used for runs to independently confirm the production results presented here. S.~D.'s contribution to this analysis was carried out  
at the Jet Propulsion Laboratory, California Institute of Technology,  
under a contract with the National Aeronautics and Space  
Administration and funded through the internal Human Resources  
Development Fund Initiative and the LISA Mission Science Office.  Some  
of the supercomputers used in this analysis were provided by funding  
from the JPL Office of the Chief Information Officer.  
  
\appendix  
\section{Waveform extrapolation}  
\label{sec:app}  
  
Here we describe the variation of Richardson extrapolation which we  
use to estimate and eliminate the largest errors arising from our  
finite difference scheme.  In Ref.\ \cite{skh07}, we showed that our  
algorithm is second order convergent.  This means that we can write  
the solution at any given resolution as  
\begin{equation}  
\Psi_c = \Psi_t + a_1 \delta r^2 + a_2 \delta \theta^2 +  
a_3 \delta r \delta \theta + {\cal O}(\delta^3)\;,  
\end{equation}  
where $\Psi_c$ is the computed solution and $\Psi_t$ is the ``true''  
solution.  The final term ${\cal O}(\delta^3)$ indicates that  
additional error terms will be third order in the grid spacing (and  
higher).  The spatial and temporal dependences of $\Psi_c$ and  
$\Psi_t$ have been suppressed.  We now perform runs at two different  
resolutions, $(\delta r_1,\delta \theta_1)$ and $(\delta r_2,\delta  
\theta_2)$, with all other parameters fixed.  The resolutions are  
chosen such that  
\begin{equation}  
\label{eq:res}  
\frac{\delta r_1}{\delta r_2} = \frac{\delta \theta_1}{\delta \theta_2} = n \;.  
\end{equation}  
Neglecting higher order terms, the two results can be written  
\begin{eqnarray}  
\label{eq:psi1}  
\Psi_{c1} & \simeq & \Psi_t + a_1 \delta r_1^2 + a_2 \delta \theta_1^2  
+ a_3 \delta r_1 \delta \theta_1 \; , \\  
\Psi_{c2} & \simeq & \Psi_t + a_1 \delta r_2^2 + a_2 \delta \theta_2^2  
+ a_3 \delta r_2 \delta \theta_2 \; .  
\end{eqnarray}  
The relation between the two resolutions, Eq.\ (\ref{eq:res}), allows  
us to write  
\begin{equation}  
\label{eq:psi2}  
\Psi_{c2}  =  \Psi_t + 1/n^2(a_1 \delta r_1^2 + a_2 \delta \theta_1^2  
+ a_3 \delta r_1 \delta \theta_1) \; .  
\end{equation}  
Subtracting Eq.\ (\ref{eq:psi2}) from Eq.\ (\ref{eq:psi1}) leaves us with  
\begin{eqnarray}  
\Psi_{c1} - \Psi_{c2} = (1 - 1/n^2)(a_1 \delta r_1^2 +  
 a_2 \delta \theta_1^2 + a_3 \delta r_1 \delta \theta_1) \; ;  
\nonumber\\  
\end{eqnarray}  
rearranging, we find  
\begin{eqnarray}  
(a_1 \delta r_1^2 + a_2 \delta \theta_1^2 + a_3  
\delta r_1 \delta \theta_1) = \frac{\Psi_{c1} - \Psi_{c2}}{1 - 1/n^2}\; .     
\label{eq:quadratic_errors}  
\end{eqnarray}  
  
To the extent that neglect of higher-order errors is warranted, this  
estimates the largest source of error.  Using Eq.\ (\ref{eq:psi1}) we  
can now estimate the ``true'' value:  
\begin{eqnarray}  
\Psi_t & \simeq & \Psi_{c1} -  (a_1 \delta r_1^2 + a_2  
\delta \theta_1^2 + a_3 \delta r_1 \delta \theta_1)  
\nonumber\\  
& = & \Psi_{c1} - \frac{\Psi_{c1} - \Psi_{c2}}{1 - 1/n^2}\;.  
\end{eqnarray}  
  
\begin{figure}[htb]  
\begin{center}  
\includegraphics[height = 65mm]{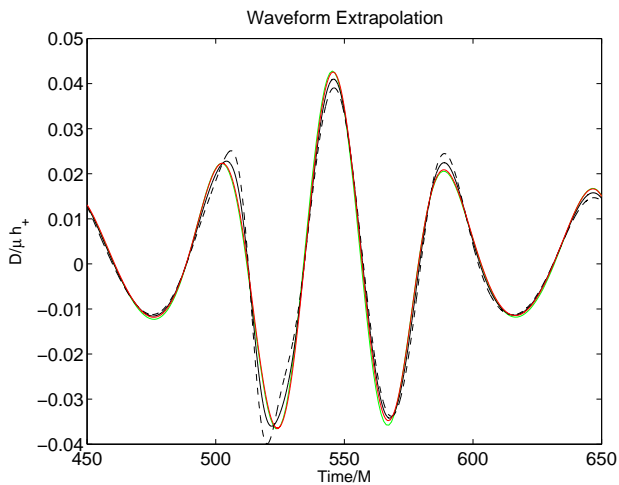}  
\caption{\label{wave_extrap} Extrapolation applied to $h_+$ for the  
$m=3$ mode from a point particle in a nearly circular geodesic with orbital  
parameters $e = 10^{-4}$, $p=6M$, and $\theta_{\rm inc} = \pi/4$ around a rotating  
black hole with spin $a/M = 0.9$. The dashed and solid black lines  
denote $h_+$ obtained with resolutions $(\delta r, \delta\theta) =  
(0.04, \pi/60)$ and $(0.026667, \pi/90)$ respectively.  The solid red  
line is the extrapolated waveform; the solid green line is the  
equivalent frequency-domain waveform.  Notice how well the  
extrapolated time-domain wave agrees with the frequency-domain result  
(which is nearly hidden by the red curve).}  
\end{center}  
\end{figure}  
  
Figure \ref{wave_extrap} illustrates the improvement that this variant  
of Richardson extrapolation can yield.  We plot $h_+$ at two different  
resolutions: $(\delta r_1, \delta\theta_1) = (0.04, \pi/60)$ and  
$(\delta r_2, \delta\theta_2) = (0.026667, \pi/90)$.  We also show the  
extrapolated waveform, and the frequency-domain prediction.  The  
particle is in a geodesic orbit with parameters $p = 6M$, $\theta_{\rm  
inc} = 45^\circ$, $e = 10^{-4}$ and the black hole has a spin of $a =  
0.9 M$.  The two time-domain calculations each differ noticeably from  
the frequency-domain result; the extrapolated waveform by contrast  
agrees very well.  This excellent agreement can be regarded as a modified three-level convergence test, whose first two levels are the time domain waveforms and third level is the frequency domain waveform. If the code were not second order convergent, our assumption for the functional form of the errors in Eq.\ (\ref{eq:quadratic_errors}) would be erroneous. This would lead to a substantial disagreement between the extrapolated and frequency domain waveforms.

\end{document}